\documentclass[12pt,dvips]{article}
\usepackage{rotating}
\usepackage{epsfig}
\usepackage{graphicx}
\usepackage{color}
\usepackage{cite}

\begin{document}

\def\thefootnote{\fnsymbol{footnote}}

\begin{flushright}
{\tt KCL-PH-TH/2012-04}, {\tt LCTS/2012-01}, {\tt CERN-PH-TH/2012-009}  \\
{\tt LBNL-, UCB-PTH-12/01} \\
{\tt ACT-1-12, MIFPA-12-01}\\
{~}\\
%{\tt arXiv:1006.NNNN} \\
%\vspace{1cm}
January 2012
%Last modified by JE on \today
\end{flushright}
%%%%%%%%%%%%%%%%%%%%%% F I G U R E %%%%%%%%%%%%%%%%%%%%%%%%%%%%%%%%%%%

%\vspace{1cm}
\begin{center}
{\bf {\Large A Historical Profile of the Higgs Boson}}
\end{center}

\medskip

\begin{center}{\large
{\bf John~Ellis}$^{a}$,
{\bf Mary~K.~Gaillard}$^{b}$ and
{\bf Dimitri~V.~Nanopoulos}$^{c}$
}
\end{center}

\begin{center}
{\em $^a$Theoretical Particle Physics and Cosmology Group, Department of
  Physics, King's~College~London, London WC2R 2LS, United Kingdom;\\
Theory Division, CERN, CH-1211 Geneva 23,
  Switzerland}\\[0.2cm]

{\em $^b$Department of Physics, University of California and
Theoretical Physics Group, \\
Bldg. 50A5104, Lawrence Berkeley National Laboratory
Berkeley, CA 94720, USA}\\[0.2cm]

{\em $^c$George P. and Cynthia W. Mitchell Institute for Fundamental Physics and Astronomy,
Texas A\&M University, College Station, TX 77843, USA;\\
Astroparticle Physics Group, Houston Advanced Research Center (HARC), Mitchell Campus, Woodlands, TX 77381, USA;\\
Academy of Athens, Division of Natural Sciences,
28 Panepistimiou Avenue, Athens 10679, Greece}\\[0.2cm]\end{center}

\bigskip

\centerline{\bf ABSTRACT}

\noindent  
The Higgs boson was postulated in 1964, and phenomenological studies of its possible
production and decays started in the early 1970s, followed by studies of its possible production in
$e^+ e^-$, ${\bar p} p$ and $pp$ collisions, in particular. Until recently, the most sensitive searches for
the Higgs boson were at LEP between 1989 and 2000, which have been complemented
by searches at the Fermilab Tevatron. The LHC has recently entered the hunt, excluding a
Higgs boson over a large range of masses and revealing a tantalizing hint in the range 119
to 125~GeV, and there are good prospects that the existence or otherwise of the Higgs boson
will soon be established. One of the most attractive possibilities is that the Higgs boson is
accompanied by supersymmetry, though composite options have yet to be excluded. 
This article reviews some of the key historical developments in
Higgs physics over the past half-century. 

\medskip
\noindent

\newpage

\section{Introduction}

The Standard Model of particle physics codifies the properties and interactions of the 
fundamental constituents of all the visible matter in the Universe. It describes successfully 
the results of myriads of accelerator experiments, some of them to a very high degree of precision.
However, the Standard Model resembles a jigsaw puzzle with one piece missing:
the Higgs boson. It may be exaggeration to term it the Holy Grail of particle physics,
and the sobriquet the `God Particle' is surely going too far, but the Higgs particle is
certainly very important. It, or something capable of replacing it, is essential for the
calculability of the Standard Model and its consistency with experimental data.

The existence of the Higgs boson was first postulated in 1964~\cite{H2}, following earlier
theoretical work that introduced spontaneous symmetry breaking into 
condensed-matter~\cite{Anderson} and particle physics~\cite{Nambu,EB,H1}. It was incorporated into the Standard Model
in 1967~\cite{Weinberg,Salam}, and shown in 1971~\cite{tHV} to lead to a calculable and predictive unified theory of the
weak and electromagnetic interactions. From 1973 onwards, with the discovery of
neutral currents~\cite{NC}, 1974 with the discovery of charm~\cite{Jpsi}, and 1983 with the discoveries of
the $W^\pm$ and $Z^0$ particles~\cite{WZ}, the predictions of the Standard Model have been
crowned with a series of successes.

Already in 1975, before the experimental discovery of charm was confirmed, we
considered that the discovery of the Higgs boson would be the culmination of the
experimental verification of the Standard Model, and we published a paper
outlining its phenomenological profile~\cite{EGN}. At the time, the Higgs boson was not on
the experimental agenda, but its star has risen over the subsequent years, first in
$e^+ e^-$ collisions~\cite{EG76} and subsequently in ${\bar p} p$ and $pp$ collisions~\cite{GGMN,GNY}, until now it is widely
(though incompletely) perceived as the primary objective of experiments at the LHC.
We anticipate that the ATLAS and CMS experiments will soon deliver their verdict on
the possible existence of the Higgs boson, providing closure on half a century of
theoretical conjecture.

In this paper we trace the trajectory of the Higgs boson from its humble theoretical
origins, through its rise to phenomenological prominence, to possible experimental
apotheosis.

\section{Prehistory}

The physicist's concept of the vacuum does not correspond to the naive idea of
`empty' space. Instead, a physicist recognizes that even in the absence of physical 
particles there are quantum effects due to `virtual' particles. For a physicist, the vacuum
is the lowest-energy state, after taking these quantum effects into account. This
lowest-energy state may not possess all the symmetries of the underlying equations
of the physical system, a phenomenon known as `spontaneous' symmetry breaking,
or `hidden' symmetry.

This mechanism of spontaneous symmetry breaking first came to prominence in
the phenomenon of superconductivity, as described in the theory of Bardeen, Cooper and Schrieffer~\cite{BCS}.
According to this theory, the photon acquires an effective mass when it propagates through certain materials
at sufficiently low temperatures, as discussed earlier by Ginzburg and Landau~\cite{GL}. In free space, 
the masslessness of the photon is guaranteed by Lorentz invariance and U(1) gauge symmetry. 
Lorentz invariance is broken explicitly by the rest frame of the superconductor, whereas the gauge
symmetry is still present, though `hidden' by the condensation of Cooper pairs of electrons~\cite{Cooper} in the lowest-energy
state (vacuum). It was explicitly shown by Anderson~\cite{Anderson} how the interactions 
with the photon of the Cooper pairs inside a superconductor caused the former to acquire an effective mass.

The idea of spontaneous symmetry breaking was introduced into particle physics by Nambu~\cite{Nambu} in 1960. 
He suggested that the low mass and low-energy interactions of pions could be understood as a reflection of a
spontaneously-broken chiral symmetry, would have been exact if the up and down quarks were massless. His
suggestion was that light quarks condense in the vacuum, much like the Cooper pairs of superconductivity.
When this happens, the `hidden' chiral symmetry causes the pions' masses to vanish, and fixes their low-energy couplings 
to protons, neutrons and each other.

\begin{figure*}[htb]
\begin{center}
%%%%%%%%%%%%%%%%%%%%%%%%%%%%%%%
\resizebox{10cm}{!}{\includegraphics{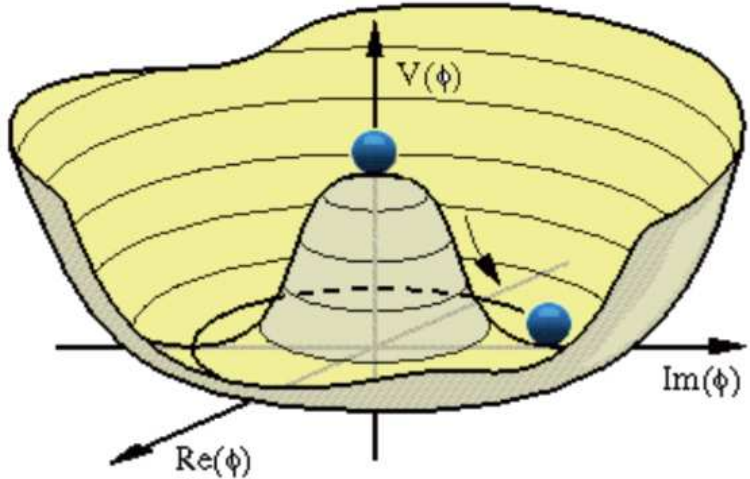}}
\end{center}
%\vspace{-3cm}
\caption{\it A prototypical effective `Mexican hat' potential that leads to `spontaneous' symmetry breaking.
The vacuum, i.e., the lowest-energy state, is described by a randomly-chosen point around the bottom of the
brim of the hat. In a `global' symmetry, movements around the bottom of the hat corresponds to a massless spin-zero
`Nambu-Goldstone' boson~\cite{Nambu,Goldstone}. 
In the case of a local (gauge) symmetry, as was pointed out by 
Englert and Brout~\cite{EB}, by Higgs~\cite{H2}
and by Guralnik, Hagen and Kibble~\cite{GHK}, 
this boson combines with a massless spin-one boson to yield a
massive spin-one particle. The Higgs boson~\cite{H2} is a 
massive spin-zero particle corresponding to quantum
fluctuations in the radial direction, oscillating between the centre and the side of the hat.
}
\label{fig:hat}
\end{figure*}

A simple model of spontaneous U(1) symmetry breaking~\cite{Goldstone} 
with a single complex field $\phi$ is illustrated in Fig.~\ref{fig:hat}.
The effective potential is unstable at the origin $\langle |\phi| \rangle = 0$. Instead, the lowest-energy state, the
vacuum, is at the bottom of the brim of the `Mexican hat', with $\langle |\phi| \rangle \ne 0$. The phase of $\phi$
is, however, not determined, and all choices are equivalent with the same energy. The system must choose
some particular value of the phase, but changing the phase would cost no energy. Hence the system
has a massless degree of freedom corresponding to rotational fluctuations of the field around the brim of the
Mexican hat. It is a general theorem proven in 1961 by Goldstone, Salam and Weinberg~\cite{GSW} that
spontaneous breaking of a global symmetry such as chiral symmetry must be accompanied by the
appearance of one or more such Nambu-Goldstone bosons. However, this is not necessarily the case if it
is a gauge symmetry that is broken, as in the non-relativistic case of superconductivity~\cite{Anderson}.
Anderson conjectured that it should be possible to extend this mechanism to the relativistic case,
as did Klein and Lee~\cite{KL}, but it was argued by Gilbert~\cite{Gilbert} that Lorentz invariance would
forbid this.

%\vspace{-2cm}

\section{And then there was Higgs}

Spontaneous breaking of gauge symmetry was introduced into particle physics in 1964 by
Englert and Brout~\cite{EB}, followed independently by Higgs~\cite{H1,H2}, and subsequently by Guralnik, Hagen and
Kibble~\cite{GHK}. They demonstrated how one could dispose simultaneously of two unwanted massless
bosons, a spinless Goldstone boson and a gauge boson of an exact local symmetry, 
combining them into a single massive vector boson in a fully relativistic theory.
The two polarizations states of a massless vector boson are combined with the single
degree of freedom of a spin-zero particle to yield the three degrees of freedom of a
massive spin-one particle.

Englert and Brout~\cite{EB} considered explicitly a non-Abelian Yang-Mills theory, assumed the formation of a vacuum
expectation value (vev) of a non-singlet scalar field, and used a diagrammatic approach to demonstrate mass
generation for the gauge field. The first paper by Higgs~\cite{H1} demonstrated that gauge symmetry
provides a loophole the `no-go' theorem of Gilbert mentioned above, and his second paper~\cite{H2} exploited this loophole
to demonstrate mass generation in the Abelian case. The subsequent paper 
by Guralnik, Hagen and Kibble~\cite{GHK} referred in its text to the Englert/Brout and Higgs papers, and
also demonstrated mass generation in the Abelian case.

The second paper by Higgs~\cite{H2} is the only one of the 1964 papers to mention
explicitly [his equation (2b)] the existence of a massive scalar particle associated with the curvature of the effective potential
that determines the vev of the charged field - see Fig.~\ref{fig:hat}. Englert and Brout~\cite{EB}
do not discuss the spectrum of physical scalars, whilst Guralnik, Hagen and Kibble~\cite{GHK} mention
a massless scalar that decoupled from the massive excitations in their model.

Over the following couple of years, there was a further important paper by Higgs~\cite{H3},
in which he discussed in detail the formulation of the spontaneously-broken Abelian theory.
In particular, he derived explicitly the Feynman rules for processes involving what has come to be
known as the massive Higgs boson, discussing its decay into 2 massive vector
bosons, vector-scalar and scalar-scalar scattering. Another important paper by Kibble~\cite{Kibble}
discussed in detail the non-Abelian case, and also mentioned the appearance of massive scalar bosons
\`a la Higgs.

The next important step was the incorporation by Weinberg~\cite{Weinberg} and by Salam~\cite{Salam} of
non-Abelian spontaneous symmetry breaking into Glashow's~\cite{Glashow} unified SU(2) $\times$ U(1) model of the 
weak and electromagnetic interactions. The paper by Weinberg was the first to observe that the scalar field
vev could also give masses to fundamental fermions.

However, the seminal papers on spontaneous breaking of gauge symmetries and
electroweak unification were largely ignored by the particle physics
community until the renormalizability of spontaneously-broken gauge theories was demonstrated by
't Hooft and Veltman~\cite{tHV}. These ideas then joined the mainstream very rapidly, thanks
in particular to a series of influential papers by B.~W.~Lee and collaborators~\cite{LZJ,FLS}.

\section{A Phenomenological Profile of the Higgs Boson}

B.~W.~Lee also carries much of the responsibility for calling the Higgs boson the Higgs boson,
mentioning repeatedly `Higgs scalar fields' in a review talk at the International Conference on High-Energy
Physics in 1972~\cite{Lee}. However, in the early 1970s there were only a few suggestions how to constrain or 
exclude the possible existence of a physical Higgs boson. One paper considered the possible effect of Higgs
exchange on neutron- and deuteron-electron scattering and derived a lower bound $m_h > 0.6$~MeV~\cite{nue}, and
another constrained Higgs emission from neutron stars, yielding the lower bound $m_h > 0.7$~MeV~\cite{SS}.
There was also a theoretical discussion of possible Higgs production in $0^+ \to 0^+$ nuclear transitions~\cite{RSW},
and its non-observation in
excited $^{16}$O and $^4$He decays led to the Higgs being excluded in the mass range $1.03~{\rm MeV} < m_H
< 18.3$~MeV~\cite{KWB}. In parallel, data on neutron-nucleus scattering were used to constrain $m_H > 15$~MeV~\cite{BE}.
The two latter were the strongest limits obtained in this period.

This was the context in which we embarked in 1975 on the first systematic study of possible Higgs phenomenology~\cite{EGN}.
Neutral currents had been discovered~\cite{NC}, and the J/$\psi$ particle~\cite{Jpsi} was thought to be charmonium, though
doubts remained and the discovery of open charm still lay in the future. The search for the intermediate vector
bosons $W^\pm$ and $Z^0$ was appearing on the experimental agenda, but the CERN ${\bar p}p$
collider that was to discover them had not yet been proposed. However, it seemed to us that the clinching
test of the spontaneous symmetry-breaking paradigm underlying the Standard Model would be discovering
the Higgs boson.

To this end, we considered the decay modes of the Higgs boson if it weighed up to 100~GeV, calculating for
the first time the loop-induced Higgs decays to photon pairs and to gluon pairs. We also estimated the cross
sections for many different mechanisms for producing the Higgs boson, intending to cover the full allowed mass range
from ${\cal O}(15)$~MeV upwards. In addition to considering the production of a relatively light Higgs boson in
hadron decays and interactions, we also considered production in $e^+ e^-$ collisions, including Higgs-strahlung 
processes such as $e^+ e^- \to Z^0 + H$~\cite{EGN}.

Back in 1975, the likelihood of a definitive search for the Higgs boson seemed somewhat remote. That was
why, rather tongue-in-cheek, we closed our paper~\cite{EGN} with the following modest words: {\it ``We should
perhaps finish our paper with an apology and a caution. We apologize to experimentalists for having no idea what is the
mass of the Higgs boson, ..., and for not being sure of its couplings to other particles, except that they are
probably all very small. For these reasons, we do not want to encourage big experimental searches for the Higgs boson,
but we do feel that people doing experiments vulnerable to the Higgs boson should know how it may turn up."}

In those early days, there was very little theoretical guidance as to the possible mass of the Higgs boson,
which is one reason why these early studies also included very low Higgs masses. One possibility that
attracted attention was that the Higgs mass was entirely due to quantum corrections, which would have yielded
$m_h \sim 10$~GeV in the absence of heavy fermions~\cite{CW}~\footnote{This was long before it was
recognized that the top quark weighed $> m_W$.}. At the other end of the mass scale, it was emphasized that the Higgs self-interactions
would become strong for $m_h \sim 1$~TeV~\cite{heavyH}.

\section{Searches for the Higgs Boson at LEP}

In addition to~\cite{EGN}, there was an early discussion of searches for the Higgs boson in $e^+ e^-$ collisions in~\cite{EG76}.
There are three important processes for producing the Higgs boson at
an $e^+ e^-$ collider: in $Z^0$ decay - $Z^0 \to H + {\bar f}f$~\cite{EG76,Bj},
in association with the $Z^0$ - $e^+ e^- \to Z^0 + H$~\cite{EGN,IK,LQT}, 
and via $W^+ W^-$ or $Z^0 Z^0$ fusion - $e^+ e^- \to {\bar \nu} H \nu, e^+ H e^-$~\cite{JP}.
The direct process $e^+ e^- \to H$ is negligible because of the small $H$ coupling
to $e^+ e^-$~\cite{EGN}, though the corresponding reaction at a muon collider, $\mu^+ \mu^- \to H$,
may be interesting~\cite{mumu}.
We also note that high-intensity lasers may be able to convert an $e^+ e^-$ collider into a
high-luminosity $\gamma \gamma$ collider, which could also be an interesting
Higgs factory~\cite{gammagamma}.

It should be noted that in the early CERN reports on LEP physics, to which~\cite{EG76}
and~\cite{LesHouches} were theoretical contributions, the concerns of our
experimental colleagues lay elsewhere. These reports contain no experimental
discussions of possible searches for the Higgs boson. The first written discussion of
which we are aware was in an unpublished 1979 report for ECFA~\cite{ECFA}, compiled by a joint working group
of theorists and experimentalists. This was followed in 1985 by a more detailed study by a
joint theoretical and experimental working group in a CERN report~\cite{LEP85} published in 1986.
Thereafter, Higgs searches were firmly in the experimental sights of the LEP collaborations, as seen
in later CERN reports on LEP physics~\cite{sights}.

\begin{figure*}[htb]
\begin{center}
%%%%%%%%%%%%%%%%%%%%%%%%%%%%%%%
\resizebox{10cm}{!}{\includegraphics{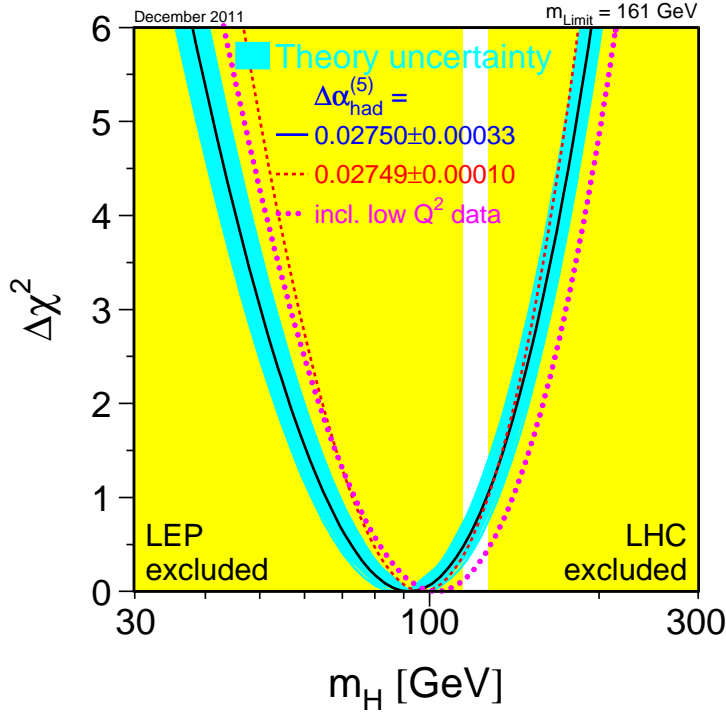}}
\end{center}
%\vspace{-3cm}
\caption{\it A compilation of information about the possible mass of the
Higgs boson~\protect\cite{LEPEWWG}. The yellow-shaded regions have been excluded by
searches at LEP~\protect\cite{LEPH}, the Tevatron collider~\protect\cite{TeVH} 
and the LHC~\protect\cite{ATLASH,CMSH}. The
black line (and the blue band) is the $\chi^2$ function for the precision electroweak data
as a function of $m_h$ (and its theoretical uncertainty). The dotted lines are obtained using
alternative treatments of the precision electroweak data~\protect\cite{LEPEWWG}.
}
\label{fig:blueband}
\end{figure*}

In parallel with the searches for the Higgs boson, notably that for
the process $Z^0 \to H + {\bar f}f$~\cite{EG76,Bj} at LEP~1 and that for
$e^+ e^- \to Z^0 + H$~\cite{EGN,IK,LQT} at LEP~2, the high-precision
electroweak data obtained at LEP, the SLC and elsewhere made it
possible for the first time to estimate the possible mass of the Higgs boson
within the framework of the Standard Model~\cite{EF}. The dominant $m_H$-dependent
corrections had been calculated earlier~\cite{Veltman}, but the possibility of using
them in conjunction with LEP data to constrain $m_H$ was not discussed
before LEP start-up, perhaps because the precision of LEP data exceeded
all previous expectations. The constraint on $m_H$ was relatively weak before
the top quark was discovered (with a mass that agreed with predictions
based on electroweak data), but the measurement of $m_t$ allowed more
accurate estimates of $m_H$ to be made. Values $< 300$~GeV were favoured
already in the early days of such studies~\cite{EF}, which have now matured to indicate
that $m_H \sim 100 \pm 30$~GeV~\cite{LEPEWWG}. This constraint is combined with the (negative)
results of the direct Higgs searches in Fig.~\ref{fig:blueband}.

The non-appearance of the Higgs boson in searches in $Z^0$ decays required $m_H > 58$~GeV~\cite{LEP1}.
Thereafter, successive increases in the LEP energy during the LEP~2 era prompted
recurrent hopes that the Higgs discovery might lie just around the corner, but instead the
lower limit on $m_H$ kept rising inexorably. Finally, in 2000 the LEP centre-of-mass
energy was pushed to 206~GeV, and a few Higgs-like events were observed,
corresponding to a mass $\sim 115$~GeV~\cite{115}. To the disappointment of many physicists,
it was not possible to push the LEP energy higher, and the difficult decision was taken
to shut LEP down at the end of the year 2000, leaving the lower limit $m_h > 114.4$~GeV
at the 95\%~CL~\cite{LEPH}. There was much speculation that this
decision forced LEP to miss out on the Higgs discovery. However, if $m_H \sim 125$~GeV 
as is now suggested by the latest LHC data~\cite{CMSH,ATLASH}, substantial extra investment in
accelerating cavities would have been necessary back in the 1990s in order to be able to push LEP to
sufficiently high energies to produce it.

\section{Searches for the Higgs Boson at Hadron Colliders}

The production of the Higgs boson at hadron colliders is more problematic than
in $e^+ e^-$ collisions. On the one hand, the backgrounds from other physical processes are large,
and on the other hand direct production via the dominant quark constituents in
the proton is small, because they have very small masses~\cite{EGN}. There is in addition, 
however, production by gluon-gluon fusion via anomalous triangle diagrams~\cite{EGN}
[see also~\cite{RU}],
as first discussed in~\cite{GGMN}. This is the dominant Higgs production mechanism at the LHC.
Another important production mechanism is Higgs-strahlung in association with a $W^\pm$ or $Z^0$, 
as was first discussed in~\cite{GNY}. This was the dominant Higgs production mechanism at the Tevatron.
A third important mechanism is $W^+ W^-$ (and $Z^0 Z^0$)
fusion, as first discussed in~\cite{CD}.

A comprehensive survey of the new physics possibilities at the Superconducting SuperCollider (SSC)
was provided in~\cite{EHLQ}, and the search for the Higgs boson naturally took pride of place. The same
was true in the survey of the new physics possibilities at the LHC provided in~\cite{EGK}, which also
included Higgs production in association with a ${\bar t} t$ pair~\cite{Zoltan}. In anticipation of Higgs 
searches at the SSC, in particular, a comprehensive survey of the theory and phenomenology of the
Higgs boson was published~\cite{HHG}. It served as the Bible for many subsequent Higgs hunters,
also at LEP and the LHC following the much-lamented demise of the SSC~\cite{Djouadi}.

After the shutdown of LEP, the lead in Higgs searches was taken by the CDF and D0 experiments,
at the Tevatron collider, where the dominant production mechanism was Higgs-strahlung in 
association with a $W^\pm$ or $Z^0$~\cite{GNY}. As the analyzed Tevatron luminosity
accummulated, CDF and D0 became able to exclude a range of Higgs masses between 156 and
177~GeV~\cite{TeVH}, as well as a range of lower masses in the range excluded by LEP. Interestingly, there
is a small excess of Higgs candidate events in a range around 130 to 140~GeV, though not
strong enough to be considered a hint, let alone significant evidence. Since the most important
Higgs decay channels for the Tevatron experiments are $H \to {\bar b}b$ and $W^+ W^-$, which
have relative;y poor mass resolution, the excess did not provide much information what value
$m_H$ might have. Unfortunately, the Tevatron was shut down in September 2011, before it could
realize its full potential for Higgs searches.

The LHC started producing collisions in late 2009, initially at low energies and starting at 7~TeV in the
centre of mass in March 2010. By the end of 2011, the ATLAS and CMS experiments had each
accumulated $\sim 5$/fb of data. Preliminary analyses of these data allowed CMS to exclude the
mass range 127 to 600~GeV~\cite{CMSH} and ATLAS to exclude three ranges: 114.4 to 115.5~GeV,
131 to 237~GeV, and 251 to 453~GeV~\cite{ATLASH}.

\section{Cornered at the LHC?}

\begin{figure*}[htb]
\begin{center}
%%%%%%%%%%%%%%%%%%%%%%%%%%%%%%%
\resizebox{10cm}{!}{\includegraphics{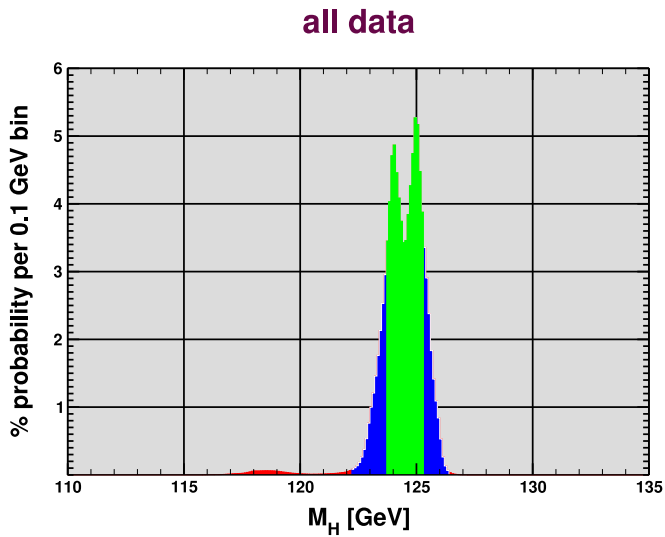}}
\end{center}
%\vspace{-3cm}
\caption{\it The probability density obtained from a global fit including
results from the direct LEP, Tevatron
and LHC searches, as well as the precision electroweak data~\protect\cite{Erler}.
}
\label{fig:Erler}
\end{figure*}

In addition to these exclusions, ATLAS and CMS have also reported on excesses of Higgs
candidate events in the non-excluded range between 115.5 and 127~GeV. In the summer of
2011 already, both ATLAS and CMS had reported a broad excess of $W^+ W^-$ events
corresponding, like the Tevatron data, to $m_H \sim 130$ to 140~GeV~\cite{Summer11}. At
the end of 2011, this was supplemented
in the ATLAS case by excesses in the decay channels $H \to \gamma \gamma$ and $ZZ$~\cite{ATLASH}, and
CMS reported smaller excesses in ${\bar b}b$ and $\tau^+ \tau^-$ as well as
$H \to \gamma \gamma$ and $ZZ$~\cite{CMSH}. The overall significance in the ATLAS experiment was
reported as 3.6~$\sigma$ and in the CMS experiment as 2.6~$\sigma$. However, whereas the
ATLAS signal is quite narrow around 126~GeV, the CMS signal is more diffuse because of the
larger relative weight of channels with poorer mass resolution, and a mass $\sim 119$~GeV
could not be excluded. At the time of writing, there is no official combination
of the ATLAS and CMS results, but one has been made unofficially, as shown in Fig.~\ref{fig:Erler}~\cite{Erler},
which is quite suggestive and quotes $m_H = 124.5 \pm 0.8$~GeV.

Since such a broad range of larger Higgs masses have been excluded by the LHC and the
Tevatron, and since a Higgs mass in this range is favoured by precision electroweak data
as well as by theoretical prejudices such as supersymmetry, it is very tempting to think
that the Higgs boson has finally been cornered. However, other options are still open
as discussed in the next section, and {\it ``extraordinary claims require extraordinary
evidence"}~\cite{Sagan}. Therefore, A signal strength of 5~$\sigma$ or more will be required before
the discovery of the Higgs boson could be announced~\footnote{Conversely, it could be
argued that proving the {\it non}-existence of the Higgs boson would be a discovery of
similar if not greater importance, so perhaps 5~$\sigma$ or more should also be required
for any {\it exclusion} of the Higgs boson.}.

Fortunately, there are good prospects for attaining this standard of proof during 2012, if
the Higgs is indeed skulking in this mass region, and the LHC accelerator continues to
work as admirably as it has so far. Current hopes are to obtain $\sim 15$/fb of
additional luminosity at a centre-of-mass energy that may be slightly higher, perhaps
8 TeV. Finally, after nearly 48 years, the capture of the Higgs boson may be imminent.

How could one be sure that any new particle discovered at the LHC really was a
Higgs boson? Various characteristics would have to be checked. First and foremost,
it would be necessary to establish that any Higgs candidate had spin zero~\cite{Hspin}. Secondly,
it should be checked whether the couplings of the candidate Higgs to other particles
were proportional to their masses (or squared masses, in the cases of the $W^\pm$
and $Z^0$). Thirdly, the magnitude of these couplings should be inversely
proportional to the expected Higgs vev, $V \sim 246$~GeV. There are good
prospects that the LHC experiments might be able to pin down the spin and check
some of the Higgs couplings~\cite{couplings}, but more detailed studies would require an $e^+ e^-$ collider -
running either at low energy in `Higgs factory' mode~\cite{ILC}~\footnote{For
an interesting option proposed recently, see~\cite{BZ}.},
or at higher energies where Higgs production would be more copious and more
Higgs couplings could be measured, including its self-couplings~\cite{CLIC} - or a $\mu^+ \mu^-$ collider~\cite{mumu}.

\section{More Higgs, Composite Higgs, Less Higgs or Higgsless?}

So far, we have focused on the least adventurous hypothesis of
a single Standard Model-like Higgs boson, but alternatives abound,
with most theorists expecting supplements to the minimal Higgs sector
of the Standard Model.

One of the simplest possibilities is that there are two complex doublets
of Higgs bosons, in which case there would be five physical Higgs bosons:
three neutral and two charged $H^\pm$. The most natural framework for such a possibility
is supersymmetry~\cite{susy}. In simple supersymmetric models the lightest of the the three neutral
Higgs boson often has couplings similar to those of the
Higgs boson in the Standard Model, whereas one of the heavier neutral
Higgs bosons would be a pseudoscalar.

The mass of the Higgs boson is linked to the magnitude of its self-coupling,
which would be fixed by supersymmetry in terms of the electroweak gauge
couplings. For this reason, supersymmetry predicts a restricted range for the mass
of the lightest Higgs boson. It used to be thought that it would necessarily weigh $< m_Z$,
but some 20 years ago it was realized that this prediction would be subject
to important radiative corrections that could push $m_H$ up to $\sim 130$~GeV
in simple supersymmetric models~\cite{ERZ}. In such a supersymmetric scenario there are no 
significant restrictions on the masses of the heavier Higgs bosons.

The range of Higgs masses favoured in simple supersymmetric models coincides with
that favoured by the precision electroweak data, and the low-mass range not yet excluded
by the LHC experiments. The hint of a Standard-Model-like Higgs boson with mass
$\sim 125$ or 119~GeV~\cite{CMSH,ATLASH} 
would fit quite nicely with supersymmetric predictions~\cite{ENOS}, and is already the
subject of many supersymmetric theoretical interpretations. 
While this scenario is theoretically very compelling, there are other possibilities, however.

An alternative to an elementary Higgs boson of the type found in supersymmetric models would be that the
spontaneous breaking of the electroweak gauge symmetry is due to a condensate in
the vacuum of pairs of new, strongly-interacting fermions~\cite{TC}, analogous to the Cooper pairs
of superconductivity~\cite{Cooper}. In this case, there would in
general be a composite scalar particle that might be accessible to experiment. However, this would not necessarily
correspond to a strongly-interacting Higgs boson, which would
have to confront issues with the precision electroweak data. On the other hand, if the composite scalar is a (relatively) light
pseudo-Goldstone boson of some higher-level broken symmetry, such as a larger chiral
symmetry~\cite{littleH} or approximate scale invariance, it would have weak interactions and could
mimic a Standard-Model-like Higgs boson to some extent.

One such example would be a pseudo-dilaton of approximate scale invariance~\cite{pseudoD}, which
would have tree-level couplings similar to those of the Higgs boson, but rescaled (and probably suppressed) by a
universal factor. The loop-induced couplings of the pseudo-dilaton to gluon and photon pairs might not share
this universal rescaling. Thus this model provides a straw person to compare with the Standard
Model Higgs scenario. It could accommodate very naturally a scalar with suppressed couplings
weighing between 127 and 600~GeV, compatible with the LHC upper limits on the production of an
intermediate-mass Higgs boson~\cite{CEO3}, 
and could even explain the observed Higgs hint at low mass. However, a Higgs-like
particle with suppressed couplings would not fulfill all the functions of the Higgs boson of the
Standard Model, e.g., in unitarizing $W^+ W^-$ scattering, and would have to be supplemented by
some other detectable degrees of freedom in the TeV energy range~\cite{G}.

Another scenario for a light Higgs-like particle is the radion~\cite{radion}, the quantum of the degree of
freedom corresponding to rescaling an extra dimension. Models with extra dimensions offer
many other possibilities, including the possibility that there is no Higgs boson at all~\cite{Higgsless}, though
such a scenario must work hard to be compatible with the precision electroweak data.

\section{Apr\`es Higgs}

The discovery~\footnote{For that is what we expect!} of the Higgs boson will mark a watershed in particle physics.
In the future, the calendar of particle physics will surely be divided into BH (before Higgs) and AH (after Higgs), with
2012 being year 0. The discovery of the Higgs will signpost the direction that both theoretical and experimental
physics will take in the decades to come.

\begin{figure*}[htb]
\begin{center}
%%%%%%%%%%%%%%%%%%%%%%%%%%%%%%%
\resizebox{10cm}{!}{\includegraphics{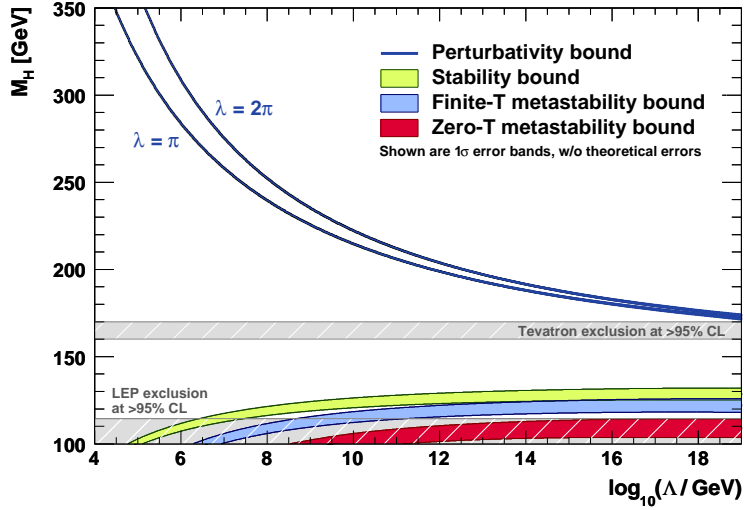}}
\end{center}
%\vspace{-3cm}
\caption{\it Bounds on the scale $\Lambda$ up to which the Standard Model may remain
valid obtained from perturbativity (solid dark blue lines) and the stability of the electroweak
vacuum under various assumptions (shaded regions)~\protect\cite{EEGHR}. Also shaded
are the regions excluded by searches at LEP and the Tevatron. LHC experiments now
exclude the range of Higgs mass where the Standard Model could remain valid up to
the Planck scale.
}
\label{fig:EEGHR}
\end{figure*}

We can be optimistic, because there are already strong arguments that, whatever the fate of the Higgs
hypothesis, there must be new physics. As we have seen, the LHC experiments already tell us that, if the
Higgs indeed exists, it must weigh either more than 600~GeV or between 115 and 127~GeV. If the Higgs
boson indeed weighs over 600~GeV, there must be new physics for two distinct reasons. One is to
reconcile such a heavy Higgs with the precision electroweak data~\footnote{But perhaps our 
interpretation of the data is at fault~\cite{MSC}? Or perhaps we are too naive in interpreting the Standard Model,
and should include the possibility of higher-dimension operators~\cite{BS}?}, and the other is to tame the blow-up
of the Higgs self-coupling, which would occur very close to the Higgs mass~\cite{EEGHR}. This is the potential
disaster indicated by the upper solid curves in Fig.~\ref{fig:EEGHR}, which represent the scale $\Lambda$ up
to which perturbative calculations in the Standard Model remain under control. Technicolour theories are
examples of such Higgs-taming theories, but they do not have an easy answer to the question of the
precision electroweak data. 

If there is something Higgs-like weighing between 127 and 530~GeV, it cannot simply have the
Standard Model couplings, or it would have been discovered already. Some mechanism must
suppress the combination of its production and decays. Perhaps it has additional decay modes
beyond the Standard Model? Perhaps its couplings are systematically suppressed, as in the
less-Higgs scenarios discussed in the previous section? Also in that case, new physics must appear~\cite{G}.

What if there is a Standard Model-like Higgs boson weighing between 115 and 127~GeV, as may be
suggested by the current indications from ATLAS and CMS~\cite{CMSH,ATLASH}? At first blush, this would sound like a
tremendous success for the Standard Model, since this coincides nicely with the range suggested by
the precision electroweak data. Indeed, but there is another problem, namely the fate of the
electroweak vacuum. If the Higgs boson is so light, quantum corrections to the self-coupling $\lambda$ of
the Higgs boson due to renormalization by loop diagrams with heavy top quark circulating
would drive $\lambda$ negative at some scale $\ll M_{Planck}$, 
triggering an instability of the present electroweak vacuum~\cite{EEGHR} - figuratively, turning down the brim of the `Mexican hat'
that was shown in Fig.~\ref{fig:hat}. This instability appears for scales $\Lambda$ in the coloured regions of Fig.~\ref{fig:EEGHR}.
It could be averted only by introducing new
physics to counter the effect of the top quark, with one very natural possibility being supersymmetry~\cite{ER}.
However, in this case, unlike the previous cases, there is no guarantee that
the necessary new physics would show up within the TeV range.

If the discovery of the Higgs boson is indeed confirmed in 2012, two new directions for physics will
open up. On the one hand, there will be a need for detailed investigation of the Higgs, to see whether
it conforms to the Standard Model paradigm or whether it exhibits deviations due to new physics.
On the other hand, the hunt will be on for whatever new physics complements the Higgs boson, be it
supersymmetry~\footnote{As we favour!}, extra dimensions, new strongly-interacting particles or ...? We look forward to the
years AH {\it (Anno Higgsi)} $> 0$.

\newpage

\section*{Acknowledgements}

The work of J.E. was supported in part by
the London Centre for Terauniverse Studies (LCTS), using funding from
the European Research Council 
via the Advanced Investigator Grant 267352. 
The work of M.K.G. was supported in part by the Director, Office of Science,
Office of High Energy and Nuclear Physics, Division of High Energy
Physics, of the U.S. Department of Energy under Contract
DE-AC02-05CH11231, in part by the National Science Foundation under
grant PHY-1002399.
The work of D.V.N. was supported in part by the
DOE grant DE-FG03-95-Er-40917.

%\newpage

\end{document}